\documentclass[prc,aps,superscriptaddress,showpacs,nofootinbib,twocolumn]{revtex4}
\usepackage{amssymb}
\usepackage[dvips]{graphicx}
\usepackage[english]{babel}
\usepackage{indentfirst}
\usepackage{amsxtra}
\usepackage{amsmath}
\usepackage{supertabular}
\usepackage{multirow}
\usepackage[mathcal]{eucal}
\usepackage[usenames]{color}
\usepackage{ulem}

\begin{document}
\title {\textbf{Beyond--mean--field effects on the symmetry energy and its slope from the low--lying dipole response of $^{68}$Ni}}
\author{M. Grasso}
\affiliation{Universit\'e Paris-Saclay, CNRS/IN2P3, IJCLab, 91405 Orsay, France}
\author{D. Gambacurta}
\affiliation{INFN-LNS, Laboratori Nazionali del Sud, 95123 Catania, Italy}
\affiliation{Extreme Light Infrastructure - Nuclear Physics (ELI-NP), Horia Hulubei National Institute for Physics and Nuclear
 Engineering, 30 Reactorului Street, RO-077125 M˘agurele, Jud. Ilfov, Romania}

\begin{abstract}
We study low--energy dipole excitations in the unstable nucleus $^{68}$Ni with the beyond--mean--field (BMF) subtracted second 
random--phase--approximation (SSRPA) model based on Skyrme interactions. First, strength distributions are compared with available experimental data and 
transition densities of some selected peaks are analyzed. The so--called isospin splitting is also discussed by studying the isoscalar/isovector character of such excitations. 

We estimate then 
in an indirect way BMF effects on the symmetry energy of infinite matter and on its slope starting from the BMF SSRPA low--lying strength 
distribution. For this, several linear correlations are used, the first one being a correlation existing between the contribution (associated 
with the low--energy strength) to the total energy--weighted sum rule (EWSR) and the slope of the symmetry energy. BMF estimates for the slope of the symmetry energy can be extracted in this way. Correlations between such a slope and the neutron--skin thickness of $^{68}$Ni and correlations between the neutron--skin thickness of $^{68}$Ni and the electric dipole polarizability times the symmetry energy are then used to deduce BMF effects on the symmetry energy. 

\end{abstract}

\pacs{ 21.60.Jz, 21.10.Re, 27.20.+n, 27.40.+z}
\maketitle

\section{Introduction}
\label{intro}

It is known that the low--lying dipole strength in neutron--rich nuclei, localized around the particle separation energy, has a strong impact on neutron radiative--capture cross sections, 
which are extremely important for nucleosynthesis processes of astrophysical interest \cite{goriely98,goriely02,goriely04,tso,lit}.  This impact was 
one of the motivations for studying 
low--energy dipole excitations, first in stable nuclei through photon scattering \cite{bart,hart}, where it was 
understood that the low--energy strength strongly depends on the neutron--to--proton ratio $N/Z$. 
Going further with the isospin asymmetry for analyzing unstable nuclei became then an important experimental challenge (see, for instance, Refs. \cite{lei,try,adrich,gibelin,wieland,rossi,martorana} and Refs. \cite{savran,lanza} for recent reviews).

In a parallel effort, several links and correlations were explored with theoretical models to relate the low--energy dipole strength of neutron--rich nuclei to the neutron--skin thickness of nuclei, to the symmetry energy of infinite matter, and to its density dependence (that is, its slope) \cite{pieka06,klim,carbone}. It is interesting to mention that various correlations (some of them will be used later) between properties of nuclei and of nuclear matter have been analyzed in the literature, starting from the observation that the existence of a neutron--skin in neutron--rich nuclei is strictly related to the density dependence of the symmetry energy \cite{typel,fur,yoshi}. 
The neutron--skin thickness was found to be correlated with the symmetry energy calculated at the saturation density $J$ as well as to its slope \cite{klim,fur,chen,cente,warda}, the slope being extremely important for example in heavy--ion collisions \cite{chen,baran,chen2,li,shetty,famiano} and in  nuclear astrophysics for the description of neutron stars \cite{horo,todd,steiner,latti}.
Correlations between the neutron--skin thickness and the product of the symmetry energy times the electric dipole polarizability were also investigated in Ref. \cite{roca2}.

It is important to stress that all such correlations were studied employing in most cases only mean--field models. For this reason, a risk could exist that these correlations  do not reflect a general behavior but simply an artifact induced by this category of models. 
 However, some of these correlations were first 
predicted within simple droplet models. This would indicate that they are probably general features of nuclei and nuclear matter and not a simple mean--field artifact. This is the case for example for the correlations found between the neutron--skin thickness of a nucleus of mass $A$ and the symmetry energy of matter $J$ minus the symmetry energy in the nucleus. The latter quantity is defined within a droplet model as $J/(1+x_A)$ where $x_A=9J/(4Q) A^{-1/3}$ and $Q$ is the so--called surface stiffness \cite{cente}. Another example is the correlation found in a droplet model between the neutron--skin thickness and $J/Q$ \cite{warda}.

To address the problem of a possible mean--field model dependence in the analysis 
of correlations between different quantities, the authors of Ref.
\cite{gulmi} made recently a study based on a Taylor expansion 
of the equation of state (EOS) of matter around the saturation density.
They analyzed in particular to what extent correlations between so--called 
low--order empirical parameters (for instance $J$ and the slope of the symmetry energy, 
entering in the lower orders of such an expansion) may be affected by uncertainties on higher--order parameters. 
To estimate uncertainties, around 50 models
(including not only relativistic and non relativistic mean--field--based models
but also many--body--perturbation--theory models with several chiral interactions) 
were used: employing all such models (not only of mean--field type)
it was shown for example that the extracted correlation coefficient between the slope 
of the symmetry energy calculated at saturation density and $J$ is quite high, $\sim$ 0.8.  

\begin{figure}
\includegraphics[scale=0.34]{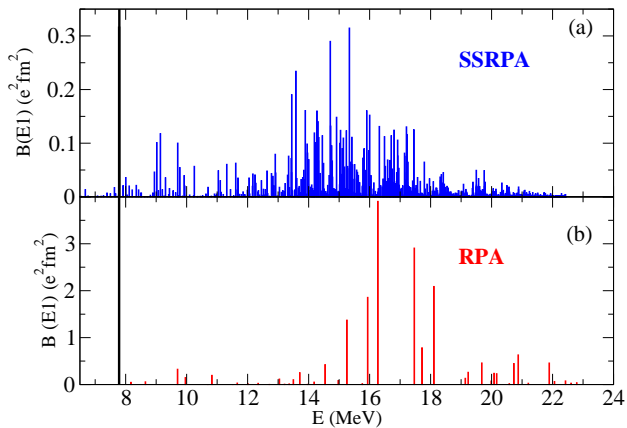}
\caption{(a) SSRPA dipole strength distribution obtained for $^{68}$Ni with the parametrization SGII; (b)  RPA dipole strength distribution obtained for $^{68}$Ni with the parametrization SGII. The vertical black line represents the neutron threshold.}
\label{dipSGIIRPA}
\end{figure}

\begin{figure}
\includegraphics[scale=0.34]{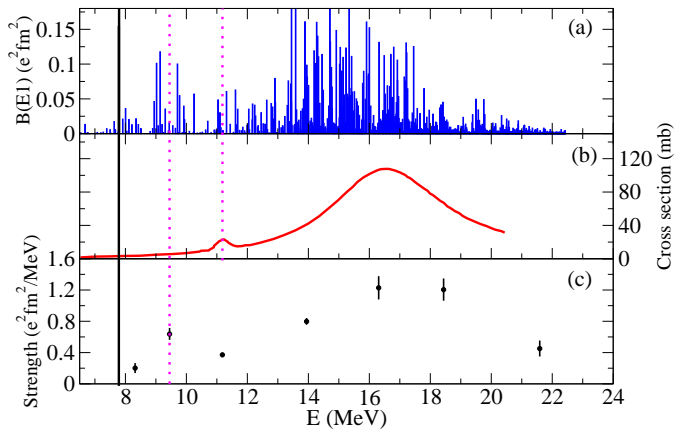}
\caption{(a) SSRPA dipole strength distribution obtained for $^{68}$Ni with the parametrization SGII; (b)  photoabsorption cross section extracted from Fig. 3 of Ref. \cite{wieland}; (c) $E1$ strength distribution extracted from Fig. 3 of Ref. \cite{rossi}. The vertical black line represents the neutron threshold. The vertical magenta dotted lines correspond to the energy values of the two experimental low--energy centroids measured in Ref. \cite{wieland} (panel (b)) and in Ref. \cite{rossi} (panel (c)). }
\label{dipSGII}
\end{figure}

It is interesting to mention that correlations between the symmetry energy and its slope were analyzed employing different experimental constraints such as those coming from heavy--ion collisions, measurements of neutron skins, electric dipole polarizabilities, masses, giant dipole resonances (GDRs), isobaric analog states, as well as costraints coming from nuclear astrophysics (see, for instance, Refs. \cite{tsang,latti1,oertel,bao,hai,latti2}).

The low--energy strength of dipole excitation spectra in neutron--rich nuclei was called 'pygmy' because of its lower energy location and of its smaller contribution to the EWSR compared to GDRs. In several cases, as indicated by the associated transition densities, such a strength was interpreted as produced by oscillations of the 
neutron skin of the nucleus against its core. This interpretation was reconsidered in some nuclei, for instance in $^{48}$Ca, where a description of the low--lying strength in terms of single--particle excitations was seen to be more appropriate \cite{ga1,ga2}. Related to this low--lying strength distribution, the mixing between isoscalar and isovector nature was also discussed \cite{burrello} as well as the mixing with complex configurations such as toroidal motions \cite{neste}.  

The first measurement of the low--lying dipole strength in the unstable nucleus $^{68}$Ni was conducted by Wieland et al. \cite{wieland} through virtual photon scattering at 600 MeV/nucleon (relativistic Coulomb excitations) at GSI. The strength was found to be centered at around 11 MeV with a contribution of 5\% to the EWSR. Later, relativistic Coulomb excitations were used once again for the same nucleus at GSI to extract the electric dipole polarizability \cite{rossi}. A slightly different result was found this time, with a centroid located at 9.55 MeV and a contribution of 2.8\% to the EWSR. The discrepancy in the value of the centroid was explained as a possible 'energy--dependent branching ratio'. 

In connection with low--lying excitations, the so--called isospin splitting was largely discuseed in the literature \cite{savran,bracco,savran2006,roca2012,savran2011,derya,crespi,pellegri,kry,crespi2015,negi}: according to this, it could be expected that the lower--energy part 
of a pygmy dipole resonance is excited by both isoscalar and isovector probes (mixed isoscalar/isovector nature) whereas the higher--energy part, 
close to the tail of the GDR, is mostly an isovector excitation. The authors of Ref. \cite{martorana} illustrated the first measurement done on $^{68}$Ni using an isoscalar probe (isoscalar $^{12}$C target) at INFN-LNS in Catania. They found a centroid placed at around 10 MeV and a contribution of 9\% to the EWSR.   

Having in the literature three slightly different experimental centroids, located at $\sim$ 11, 9.55, and 10 MeV, we focus in this work on the study of the low--lying dipole response of the unstable nucleus $^{68}$Ni using the BMF SSRPA model with Skyrme interactions \cite{ga2015}. Section \ref{ssrpapre} presents our predictions obtained with the Skyrme parametrization SGII \cite{sgii1,sgii2} and the comparison with the available experimental results. 
Some selected transition densities are analyzed and a possible isospin splitting is discussed by studying the isoscalar/isovector character of these excitations.

We then move to the symmetry energy and its density dependence, which is characterized by its slope. Section \ref{JLeffe} illustrates BMF effects on the symmetry energy and its slope, using as a starting point correlations existing between the percentage of the EWSR associated with the low--lying dipole strength in $^{68}$Ni and the slope of the symmetry energy (Subsec. \ref{EW}). Correlations between the neutron--skin thickness and the slope of the symmetry energy (Subsec. \ref{delta}) and correlations between the electric dipole polarizability times the symmetry energy and the neutron--skin thickness (Subsec. \ref{alpha}) are then used to estimate the impact of BMF calculations on the symmetry energy and its slope (Subsec. \ref{bmfinal}). 
This estimation is of course qualitative and we are not going to provide any quantitative predictions. 
Our aim is to show to what extent such quantities can be modified by effects induced by beyond--mean--field
correlations.
Conclusions are drawn in Sec. \ref{conclu}.   

\section{SSRPA predictions and comparison with measurements for the low--lying strength of $^{68}$Ni}
\label{ssrpapre}
The SSRPA model is applied to compute the dipole strength distribution of $^{68}$Ni with the Skyrme parametrization SGII.
A cutoff of 80 (50) MeV 
 is chosen for the 1p1h (2p2h) sector and the diagonal approximation is adopted in the 2p2h matrix used 
 for the calculation of the corrective term in the subtraction procedure \cite{ga2015}. 
 Figure \ref{dipSGIIRPA} represents the SSRPA results (panel(a))  for the transition operator
 \begin{equation}
  \label{op}
  T_{1M}= \frac{Z}{A}\sum_{n=1,N} r_n Y_{1M} (\hat{r}_n)-\frac{N}{A}\sum_{p=1,Z} r_p Y_{1M}(\hat{r}_p)
 \end{equation}
and the comparison with the random--phase--approximation (RPA) spectrum (panel(b)) computed with the same Skyrme functional.
We applied the procedure described in Ref. \cite{roca2012} in order to project out possible admixtures with spurious
components. We found that the this procedure affects the transition probability and the EWSR percentage by less than 0.01 \%, showing that our self--consistent approach is reliable. 
A vertical line located at 7.792 MeV indicates the neutron threshold.

Before describing the low--lying spectrum (below $\sim$ 12 MeV), which is the focus of this work, we may compare the strength distributions in the GDR region and say that, as expected, the SSRPA spectrum is much denser in this region compared to the RPA case, describing the physical fragmentation and width of the resonance.

 Let us focus on the region below $\sim 12$ MeV, to identify there BMF effects. We first observe that there is more strength in the SSRPA spectrum which results, as we will see later, in a higher percentage of EWSR (in the BMF model) computed up to a low--energy cutoff, which separates the pygmy excitation and the GDR. This difference reflects a BMF effect. We also observe that the SSRPA low--energy distribution shows peaks concentrated around 9 and 10 MeV, as well as several peaks located just above 11 MeV, which means that a non negligible strength is predicted by the SSRPA model in the regions where thet three experimental centroids are located. On the other side, in the low--energy part of the RPA spectrum there are much less peaks, the highest one being located between 9.5 and 10 MeV. Another isolated peak is placed below 11 MeV and there is practically no strength above 11 MeV. We may conlude that BMF effects are important to provide a larger fragmentation of the strength leading to a better coverage, in SSRPA, of the region where the three experimental centroids were found. 
Looking at Fig. \ref{dipSGIIRPA}, one observes a kind of separation at $\sim$ 12 MeV between the low--energy strength and the strength that may be associated with the 
tail of the giant resonance (we will discuss this later). Since we are going to estimate BMF effects related to the low--energy strength computed with the interaction SGII, we calculate the percentage of the EWSR up to 12 MeV. Such a percentage is equal to 3.75 (whereas it is equal to 2.35 within the SGII--RPA model). For reasons of consistency, we are going in what follows to compute all the needed percentages of EWSR up to 12 MeV.  
We recall that the experimental low--energy contribution to the EWSR was found to be equal to 5, 2.6, and 9 \% 
in the three experiments mentioned here.

Figure \ref{dipSGII} shows the comparison of SSRPA results with the experimental data. The comparison with the experimental data is carried out by comparing only the location of the energy peaks of the measured and predicted excitation spectra (vertical axes describe different quantities in the three panels). 
Panel (a) shows the SSRPA $B(E1)$ distribution. Panels (b) and (c) are extracted, respectively, from the upper panel of Fig. 3 of Ref. \cite{wieland} and from Fig. 3 of Ref. \cite{rossi}. They represent, respectively, the photoabsorption cross section of Ref. \cite{wieland} and 
the $E1$ strength distribution of Ref. \cite{rossi}.  For the discussion, 
we remind once again that the low--energy centroid found in Ref. \cite{martorana} is placed at 10 MeV, whereas the centroids of Refs. \cite{wieland} and \cite{rossi} are located at  $\sim$ 11 and at 9.55 MeV, respectively, as Fig. \ref{dipSGII} indicates.

\begin{widetext}

\begin{figure}
\includegraphics[scale=0.5]{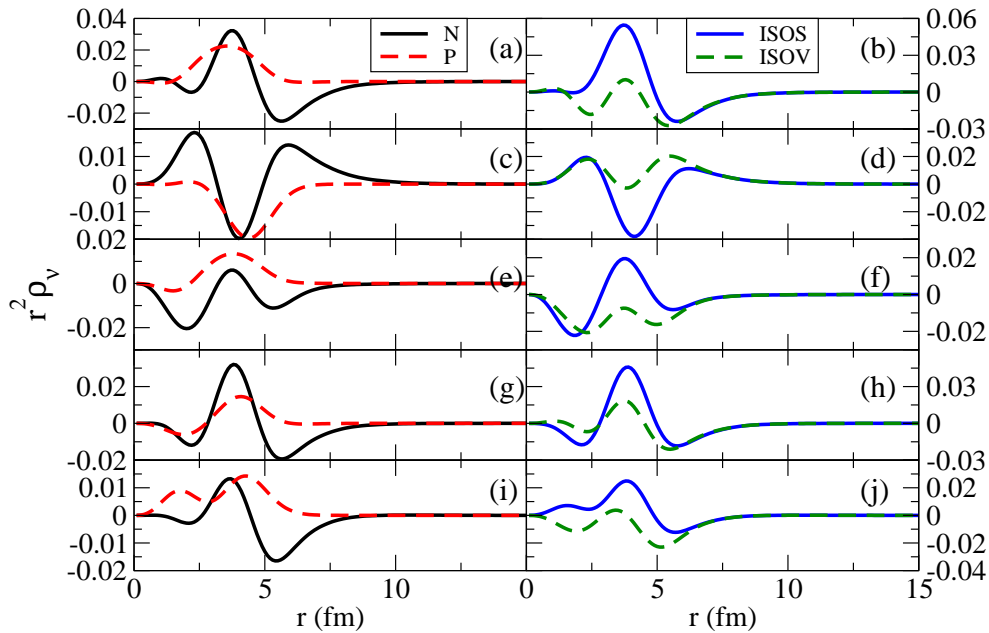}
\caption{Neutron and proton transition densities: panels (a), (c), (e), (g), and (i), and isoscalar and isovector transition densities: panels (b), (d), (f), (h), and (j). Transition densities are calculated for the peaks shown on Fig. \ref{dipSGII} and located at 9.14 ((a) and (b)), 9.70 ((c) and (d)), 10.25 ((e) and (f)), 11.10 ((g) and (h)), and 11.31 ((i) and (j)) MeV. The used Skyrme parametrization is SGII. The transition densities are multiplied by $r^2$ and are thus expressed in units of (fm$^{-1}$).   }
\label{TDSGII}
\end{figure}

\end{widetext}

Transition densities are shown in Fig. \ref{TDSGII} for two peaks in the region around 9.55 MeV (namely, the peaks located at 9.14 (panels (a) and (b)) and 9.70 (panels (c) and (d)) MeV), for one peak in the region around 10 MeV (namely, the peak located at 10.25 MeV  (panels (e) and (f)), and for two peaks in the region around 11 MeV (namely, the peaks located at 11.10 (panels (g) and (h)) and 11.31 (panels (i) and (j)) MeV). Neutron/proton and isoscalar/isovector transition densities are shown on panels (a), (c), (e), (g), (i) and (b), (d), (f), (h), (j), respectively.

\begin{figure}
\includegraphics[scale=0.4]{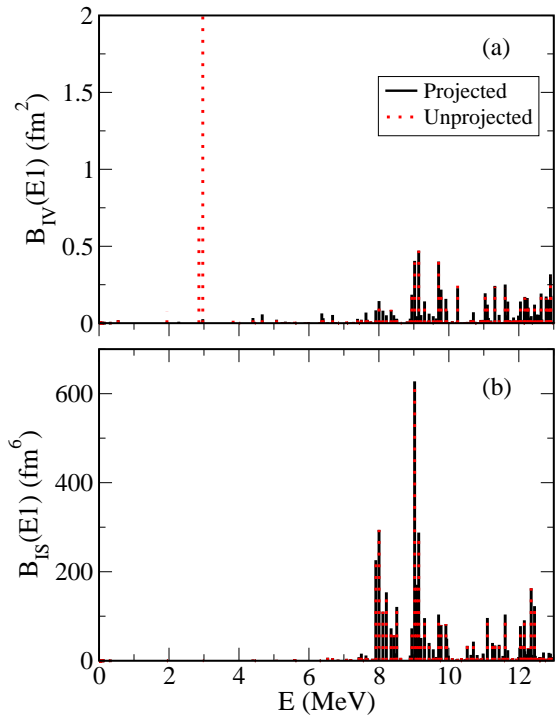}
\caption{ Isovector (a) and isoscalar (b) low--lying strength 
distributions, corresponding respectively
to the operators (\ref{op_iv}) and (\ref{op_is}). Full lines represent the 
projected results whereas the dotted lines are obtained without correcting for spurious 
components.}
\label{fig:strength}
\end{figure}

By looking at the neutron and proton transition densities (left panels) one 
observes a systematic dominant neutron contribution located at the surface of 
the nucleus, which is a typical feature of a pygmy excitation in its 
conventional interpretation. On the other side, by looking at the isovector and 
isoscalar transition densities (right panels) one does not observe any specific 
well--defined evolution going from upper to lower panels (increasing excitation 
energy). In 
other words, one does not observe any clear isospin 
splitting, which would imply a mixed isoscalar/isovector nature in the lower 
energy region and a dominant isovector nature in the higher energy part of the 
low--lying spectrum. However, before concluding on this aspect and 
in order to have a more quantitative insight on the  isospin nature  of the low--lying excitations,
we plot in Fig. \ref{fig:strength} the transition probabilities associated with the isovector 
 \begin{equation}
  \label{op_iv}
  T_{1M}^{IV}= \sum_{n=1,N} r_n Y_{1M} (\hat{r}_n)-\sum_{p=1,Z} r_p Y_{1M}(\hat{r}_p)
 \end{equation}
and the isoscalar 

 \begin{equation}
  \label{op_is}
  T_{1M}^{IS}= \sum_{i=1,A} (r_i^3 - \frac{5}{3} \langle r^2 \rangle r_i )Y_{1M} (\hat{r}_i)
 \end{equation}
 dipole operators. Both unprojected and projected (that is, with spurious--component corrections) 
results are shown. 
 The comparison between the isovector (panel (a)) and the isoscalar (panel (b)) transition strengths
 shows  a very strong isospin mixing that might explain that  
different states may be excited with different probes.  

Whereas in the isovector distribution the strength values are more or less comparable among themselves in the whole 
energy window, from $\sim$ 7 to $\sim$ 13 MeV, in the isoscalar distribution one observes that the strength is more important in the lower--energy part (below $\sim$ 10 MeV) than in the higher--energy part of the spectrum. This indicates the presence of an isoscalar/isovector splitting \cite{savran2006,endres}. Similar conclusions may be drawn by comparing the isoscalar strength with the electromagnetic one shown in Figs. \ref{dipSGIIRPA} and \ref{dipSGII}. 

From the figure we can also clearly see how
the effect  of the projection of the spurious components is acting almost exclusively
on the lowest states located at around 3 MeV, corresponding to the spurious mode.

\section{BMF effects on the symmetry energy and its slope}
\label{JLeffe}

Let us first write the expressions of the quantities that we are going to use in what follows. The neutron--skin thickness of a nucleus is defined as the difference between the neutron and 
the proton root--mean--square radii, 
\begin{equation}
\Delta r_{np} = <r^2>^{1/2}_n - <r^2>^{1/2}_p.
\label{rms}
\end{equation}

By introducing the isospin--asymmetry parameter $\delta=(\rho_n-\rho_p)/\rho$, where $\rho_n$, $\rho_p$, and $\rho$ are the neutron, proton, and total densities, respectively, one can write the EOS for asymmetric matter as 
\begin{equation}
E (\rho,\delta)=E(\rho, \delta=0) + S(\rho) \delta^2 + \mathcal{O}(\delta^4), 
\label{eos}
\end{equation}
where $S(\rho)$ is called symmetry--energy coefficient, 
\begin{equation}
S(\rho)=\frac{1}{2} \frac{\partial^2}{\partial \delta^2} E(\rho,\delta) |_{\delta=0}.
\label{sym}
\end{equation}

By truncating Eq. (\ref{eos}) at the quadratic term (parabolic approximation) the symmetry--energy coefficient may be computed as the difference between the EOSs of neutron and symmetric matter.  
 The value of $S(\rho)$ at the saturation density $\rho_0$ is often called $J$, $J=S(\rho_0)$. 
One can expand $S(\rho)$ around the saturation density, 
\begin{equation}
S(\rho)=J+L \frac{\rho-\rho_0}{3 \rho_0} +\frac{1}{2} k_{sym} (\frac{\rho-\rho_0}{3 \rho_0})^2+\mathcal{O} [(\rho-\rho_0)^3],
\label{expaS}
\end{equation}
where $L$ and $k_{sym}$ are related to the first and second derivatives of $S(\rho)$, respectively. In particular, $L$ is the slope of 
the symmetry energy,
\begin{equation}
L=3\rho_0\frac{\partial S(\rho)}{\partial \rho} |_{\rho=\rho_0}. 
\label{slope}
\end{equation}

To visualize the linear correlations that we are going to employ, we have chosen four Skyrme parametrizations having quite different values of $L$, namely, 
SGII, SIII \cite{siii}, SkI3 \cite{rei1}, and SkI4 \cite{rei1}.
The associated values of $J$ and $L$ are reported on Table \ref{symL}. There, the values of $J$ and $L$ for SGII and  SIII are extracted from 
Ref. \cite{warda} whereas the values of $J$ and $L$ for SkI3 and SkI4 are extracted from Ref. \cite{rei2}. 

\begin {table} 
\begin{center}
\begin{tabular}{ccc}

                     \hline
\hline
   Skyrme    &  $J$ (MeV) & $L$ (MeV) \\
\hline
SIII  & 28.16 & 9.90 \\
SGII & 26.83 & 37.70 \\
SkI4 & 29.50 & 60.00 \\
SkI3 & 34.27 & 100.49 \\
\hline
\hline
\end{tabular}
\end{center}
\caption{Mean--field values for the symmetry--energy coefficient computed at the saturation density $J$ and for the associated slope $L$ for the four Skyrme parametrizations indicated on the first column.}
\label{symL}
\end {table} 
  
\begin{figure}
\includegraphics[scale=0.36]{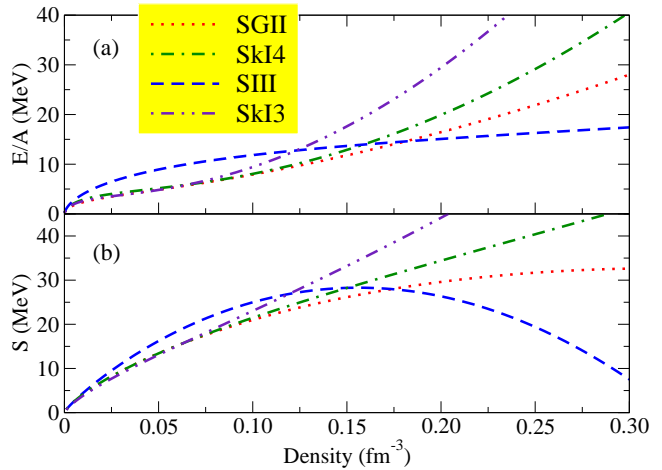}
\caption{(a) Mean--field EOSs of pure neutron matter computed with the four Skyrme parametrizations SIII, SGII, SkI3, and SkI4. (b) Mean--field symmetry--energy coefficient associated with the four Skyrme parametrizations used in (a).}
\label{nmeos}
\end{figure}

 Such $J$ and $L$ values are associated with EOSs computed at the mean--field level, corresponding to the leading (first) order of the Dyson equation. The different values of $L$ are produced by quite different mean--field EOSs for pure neutron matter, as displayed in Fig. \ref{nmeos}, where mean--field EOSs for neutron matter and the corresponding mean--field symmetry--energy coefficients are plotted for the four Skyrme parametrizations. We are aware of the bad quality of the neutron matter EOS produced by the parametrization SIII, for which the value of the slope $L$ is indeed very low, but we are going to use this case only to visualize linear correlations and perform linear fits. In any case, results and predictions will be discussed only for the parametrization SGII which represents an illustrative reasonable case. 

It is expected that, when BMF models are used, the corresponding symmetry energies and slopes evolve because they have to be associated now with BMF EOSs for infinite matter.  
We estimate here such a BMF effect in an indirect way, by the analysis of the BMF SSRPA low--energy dipole strength. In particular, we employ for this several correlations that have been analyzed in the literature. We stress that we are not going to provide any precise predictions. Our aim is to estimate qualitatively to what extent a BMF model can have an impact on such quantities.

\subsection{EWSR and slope of the symmetry energy $L$}
\label{EW}
\begin{figure}
\includegraphics[scale=0.36]{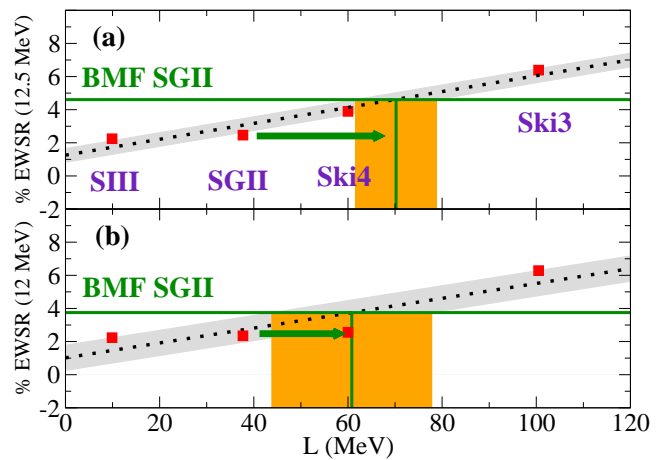}
\caption{(a) Percentage of the EWSR calculated up to 12.5 MeV for the nucleus $^{68}$Ni with the RPA model with the four Skyrme parametrizations SIII, SGII, SkI3, and SkI4 (red squares) as a function of the slope 
$L$ (mean--field values) of the symmetry energy. A linear fit is carried out on the four points (dotted line). The grey area represents the associated  uncertainty band (see text). The BMF SSRPA \% EWSR value is included for the case SGII (horizontal green line) and a BMF SGII value of $L$ is correspondingly extracted (vertical green line) with an associated uncertainty (orange area). (b) Same as in (a), where the EWSR percentage is computed up to 12 MeV. }
\label{SR}
\end{figure}

We use first correlations found between the percentage of the EWSR associated with the low--lying dipole strength and the slope $L$ \cite{klim,carbone}. We carry out RPA calculations with the four Skyrme parametrizations of Table \ref{symL}. The contribution (percentage) to the EWSR is evaluated
up to an energy of 12 MeV. The total EWSR is satisified better than 1\%. The corresponding points are shown on Fig. \ref{SR}(b) as a function of the slope $L$ (mean--field values of $L$, taken from Table \ref{symL}). We perform a linear fit (black dotted line in the figure) and associate an uncertainty band which is evaluated in the following way. 
Using the linear fit done on the points obtained with the RPA EWSR percentages and the mean--field $L$ values for infinite matter, we are going to extract a BMF value of $L$ for a given BMF SSRPA percentage of the EWSR. To construct an uncertainty band we compute, for each RPA percentage of the EWSR, the horizontal distance between the corresponding mean--field value of $L$ and the point located on the linear--fit curve. We take the average of the distances computed in this way for the four interactions and we define as uncertainty a band having an horizontal width equal to twice such an average.  
This is represented by the grey area in the figure. 
For the  Skyrme parametrization SGII we report on the linear fit the corresponding prediction for the SSRPA percentage of the EWSR (computed up to 12 MeV). Such a percentage, equal to 3.75, is larger than the RPA value for SGII. Using the linear fit and the uncertainty band, we may then extract a BMF value of $L$, with an associated uncertainty. 

Whereas the mean--field value of $L$ is 37.7 MeV for the parametrization SGII, the extracted BMF value is 60.815 $ \pm$ 16.982 MeV. The slope of the symmetry energy was thus increased owing to BMF effects, which implies that BMF effects tend to produce a stiffer EOS for pure neutron matter. 

We carried out the same analysis with a cut at 12.5 MeV. This is shown on Fig.  \ref{SR}(a). The percentages of EWSR are obviously increased for both RPA and SSRPA calculations compared to the previous case. One may extract in the same way a BMF value for $L$. This value is larger than the one associated with a cut at 12 MeV, as the figure shows. However, the corresponding uncertainty is more narrow. One may thus conclude that, using the BMF estimation obtained at 12 MeV (which  seems to be anyway the most reasonable choice looking at the strength distributions), the associated uncertainty band is sufficiently large to include BMF estimations that may be obtained by slightly increasing the cut.

\begin{figure}
\includegraphics[scale=0.34]{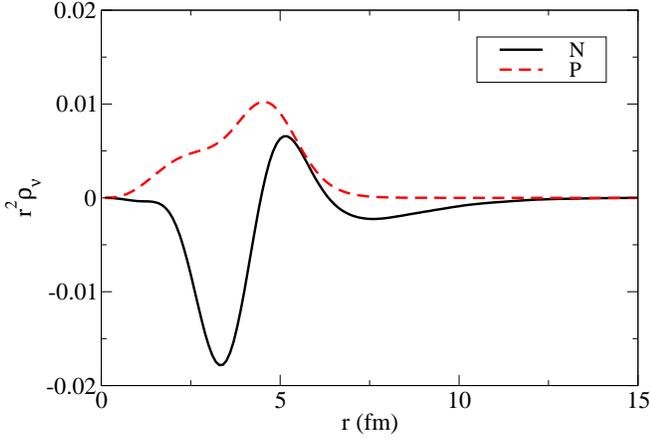}
\caption{ Neutron and proton transition densities for the state located at 12.17 MeV. The transition densities are multiplied by $r^2$ and are thus expressed in units of (fm$^{-1}$). }
\label{transi}
\end{figure}

We have indeed checked by looking at the transition densities that a cut at 12 MeV is the best choice. Below 12 MeV, the neutron and proton transition densities are similar to the cases shown on Fig. \ref{TDSGII}, where typical features of a pygmy resonance may be recognized. However, above 12 MeV, the shapes of the densities indicate that a transition towards the tail of the GDR starts to occur. An illustrative example is shown in Fig. \ref{transi}, where neutron and proton transition densities for the state located at 12.17 MeV are presented. Clear features of a pygmy excitation are missing there and a more mixed behavior starts to appear.   

By including more functionals to describe the linear correlation that we employ here, a higher spreading of the results around the linear fit would of course result. However, in this work, the scope is not a precise indication. By using such a linear correlation and, more importantly, by using the fact that the percentage of the EWSR associated with the low--energy part of the spectrum is larger in SSRPA than in RPA for each given functional, we extract a qualitative estimation. This indicates to what extent one may expect that such a BMF result can affect $L$ and modify it from its mean--field value.

\begin{figure}
\includegraphics[scale=0.36]{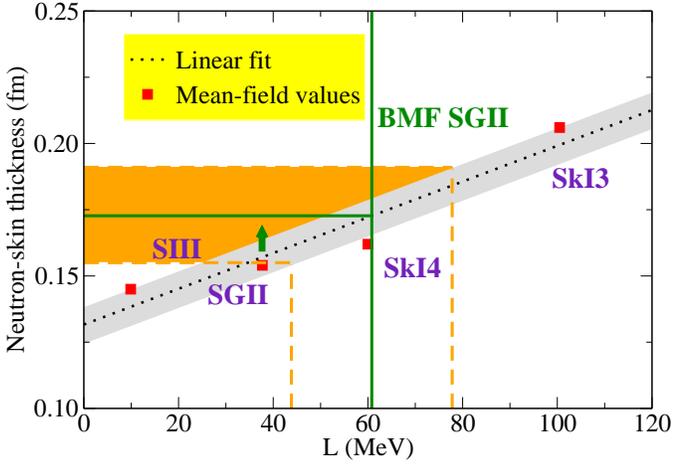}
\caption{Neutron--skin thickness computed for the nucleus $^{68}$Ni with mean--field calculations using the four Skyrme parametrizations SIII, SGII, SkI3, and SkI4 as a function of the slope 
$L$ (mean--field values) of the symmetry energy (red squares). A linear fit is carried out on the four points (dotted line) and an uncertainty band (grey area) is also displayed (see text). The BMF $L$ value for SGII is included (vertical green line) with its uncertainty band which is represented by vertical orange dashed lines, and the corresponding BMF value for $\Delta r_{np}$ is extracted (horizontal green line) with an associated uncertainty (orange area).}
\label{skinl}
\end{figure}

\begin{figure}
\includegraphics[scale=0.36]{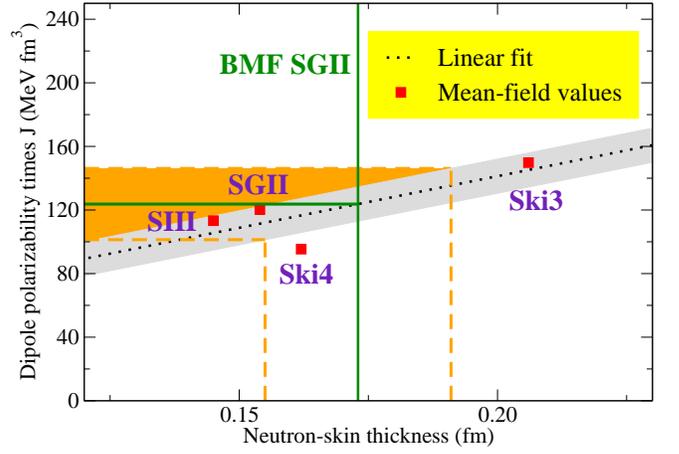}
\caption{Electric dipole polarizability computed with the RPA model for the nucleus $^{68}$Ni times the mean--field value of $J$ versus the mean--field value of $\Delta r_{np}$, using the four Skyrme parametrizations SIII, SGII, SkI3, and SkI4 (red squares). A linear fit is carried out on the four points (dotted line) and an uncertainty band (grey area) is also displayed (see text). The BMF value for $\Delta r_{np}$ is included for SGII (vertical green line) with its uncertainty band represented by vertical orange dashed lines. The corresponding BMF value of $\alpha J$ is extracted (horizontal green line) with its uncertainty (orange area). Since $\alpha$ is the same in RPA and in SSRPA, this procedure provides a BMF value for $J$. }
\label{alphaJL}
\end{figure}

\begin{figure}
\includegraphics[scale=0.36]{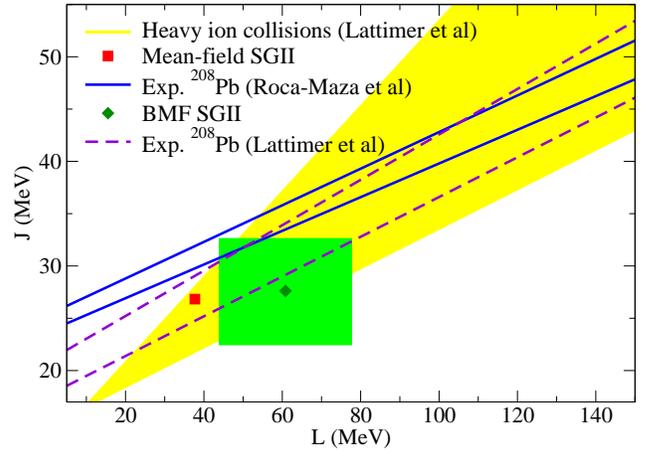} 
\caption{Experimental constraints coming from the measurement of the electric dipole polarizability in $^{208}$Pb extracted from Refs. \cite{roca2} (region between the blue solid lines) and \cite{latti2} (region between the indigo dashed lines). Experimental constraints provided by heavy--ion collisions are also reported (yellow area), which are extracted from Ref. \cite{latti2}.  The mean--field and the BMF points are indicated for the Skyrme parametrization SGII as a red square and a dark green diamond, respectively. The uncertainty associated with the BMF value is also reported as a light green area.}
\label{JL}
\end{figure}

\subsection{Neutron--skin thickness $\Delta r_{np}$ and $L$}
\label{delta}
We use now the correlations existing between the neutron--skin thickness and the slope $L$. Such correlations have been analyzed for example in Refs. \cite{cente,warda}. We perform Hartree-Fock calculations for the nucleus $^{68}$Ni with the four Skyrme parametrizations that we have chosen and report on Fig. \ref{skinl} the corresponding values for $\Delta r_{np}$ as a function of $L$ (mean--field values). 
Again, we perform a linear fit (black dotted line) and estimate in the same way as done before an uncertainty band. 
This time, we are going to extract a BMF $\Delta r_{np}$ value by using the BMF $L$ value estimated in the previous step.  
 For each mean--field value of $L$ associated with a given Skyrme parametrization we compute the vertical distance between the Hartree-Fock neutron--skin thickness and the corresponding point on the linear--fit curve. We make an average of the four distances and construct in this way an uncertainty band as a band having a vertical width equal to twice this average (grey area in the figure). 
We can now report on the figure the SGII  BMF value of $L$ (with its uncertainty represented by the two vertical orange dashed lines) extracted in the previous step. We produce in this way an estimation for a BMF value of the neutron--skin thickness with an associated uncertainty (orange area). 
We mention that the two steps illustrated in Subsecs. \ref{EW} and \ref{delta} were employed in Ref. \cite{carbone} as a way to extract constraints 
on the neutron--skin thickness by the analysis of low--lying pygmy resonances. We carry out here the same procedure to extract a BMF estimation of $\Delta r_{np}$. 

It is interesting to see that the 
neutron--skin thickness of the nucleus is impacted by BMF effects. The mean--field value for $\Delta r_{np}$ is 0.154 fm with the parametrization SGII, whereas the BMF value is equal to 0.173 $\pm$ 0.018 fm with the same interaction. The mean--field value is located at the lower border of the BMF uncertainty band.
 Qualitatively, one may conclude that BMF effects tend to increase the neutron skin of the nucleus.   

\subsection{Electric dipole polarizability times $J$ and $\Delta r_{np}$}
\label{alpha}

The authors of Ref. \cite{roca2} discussed a linear correlation existing between the electric dipole polarizability $\alpha$ times $J$ and the neutron--skin thickness of a given nucleus. We use such a correlation to extract a BMF value of $J$ from the BMF value of $\Delta r_{np}$ which was estimated in Subsec. \ref{delta}. By construction, the dipole polarizability $\alpha$ is the same in RPA and in SSRPA (owing to the subtraction procedure \cite{ga2015}). Thus, we compute $\alpha$ for $^{68}$Ni within the RPA model and obtain $\alpha=$ 4.48 fm$^3$ for the parametrization SGII. This value will be used also for the SSRPA model. Figure \ref{alphaJL} shows the correlation between $\alpha J$ and $\Delta r_{np}$ using mean--field--based calculations done with the four Skyrme parametrizations SIII, SGII, SkI3, and SkI4. A linear fit is carried out again (dotted line) and an uncertainty band is estimated. Here, we are going to use a BMF value for $\Delta r_{np}$ to extract a BMF value for $\alpha J$. We evaluate for the four points the vertical distance between the mean--field values of $\alpha J$ 
and the linear--fit curve. A vertical uncertainty is defined as twice the average of the four distances. The BMF value of $\Delta r_{np}$ with its 
uncertainty band is reported and a BMF value of $J$ may be extracted ($\alpha$ being the same as in RPA), equal to 27.617 $\pm$ 5.004 MeV. 
The symmetry energy at saturation density was slightly increased by BMF effects 
even if one may note that the mean--field value, 26.83 MeV, falls inside the uncertainty 
band associated with the BMF estimation.

\subsection{BMF values for $J$ and $L$}
\label{bmfinal}

The correlation between the electric dipole polarizability times the symmetry energy and the neutron skin thickness discussed in Subsec. \ref{alpha} was extended to a correlation between the dipole polarizability times the symmetry energy and the slope $L$ in Ref. \cite{roca2013}.  This was done in particular for the nucleus $^{208}$Pb. Using the experimental measurement of the dipole polarizability, a relation between $J$ and $L$ was then extracted in Ref. \cite{roca2}. Based on the experimental value of (19.6 $\pm$ 0.6) fm$^3$ such a relation is shown on Fig. \ref{JL} as the region between the two blue solid lines.  

On the other hand, using the recent value of (20.1 $\pm$ 0.6) fm$^3$ reported by Tamii et al. \cite{pb}, Lattimer and Steiner extracted a slightly different constraint for $J$ and $L$ in Ref. \cite{latti2} which is displayed in Fig. \ref{JL} as the region between the indigo dashed lines. 
We chose this case of $^{208}$Pb  as an illustrative example to show that such empirical constraints have indeed to be taken as qualitative indications: sligthly different measured values of the dipole polarizability may modify the empirical region which  constrains the values of $J$ and $L$.   
We have also extracted from Ref. \cite{latti2} the empirical constraint on $J$ and $L$ provided by heav--ion collisions (yellow area). 

 On the same figure, the mean--field value of $J$ and $L$ corresponding to the parametrization SGII is included together with the BMF estimation and the associated uncertainty area.  

We observe that the value of $L$ is much more strongly impacted by BMF effects than the value of $J$. 
The mean--field point is located outside the region defined by the blue solid lines 
 (extracted from Roca-Maza et al \cite{roca2}), whereas the BMF area is compatible with this region if the uncertainty band is taken into account.  
 The two points (mean--field and BMF) are both compatible with the empirical constraint provided by the indigo dashed lines (extracted from Lattimer and Steiner \cite{latti2}).  The mean--field point is located inside this area whereas the BMF one is compatible with it if one considers the uncertainty region. The yellow band in the figure represents the empirical constraint provided by heavy--ion collisions and also extracted from Ref. \cite{latti2}. The mean--field point is placed at the  border of this area, whereas the BMF case tends to favour a higher slope $L$, which implies a stiffer EOS for pure neutron matter. 
Mean--field values and the corresponding BMF estimations for $L$, $\Delta r_{np}$, and $J$ are 
summarized on Table \ref{summa} for the parametrization SGII. 

\begin {table} 
\begin{center}
\begin{tabular}{cccc}

                     \hline
\hline
 &   $L$ (MeV)    &  $\Delta r_{np}$ (fm) & $J$ (MeV) \\
\hline
Mean field & 37.70 & 0.154 & 26.83 \\
BMF & 60.815 $\pm$ 16.982    &  0.173 $\pm$ 0.018 & 27.617 $\pm$ 5.004   \\
\hline
\hline
\end{tabular}
\end{center}
\caption{Mean--field values and BMF estimations obtained with the parametrization SGII for the slope of the symmetry--energy $L$, the neutron--skin thickness $\Delta r_{np}$, and the symmetry energy calculated at the saturation density  $J$.}
\label{summa}
\end {table}

\section{Conclusions}
\label{conclu}

In this article we have studied the low--energy dipole strength distribution of the unstable nucleus $^{68}$Ni with the BMF SSRPA model based on Skyrme interactions. The parametrization SGII is chosen as illustrative case. 
First, the low--energy response is compared with three available experimental measurements, which led to the experimental centroids of 11 \cite{wieland}, 9.55 \cite{rossi}, and 10 \cite{martorana} MeV. The SSRPA model provides peaks around these three energy values. 
Transition densities were analyzed for two peaks in the region of 9.55 MeV, one peak around 10 MeV, and two peaks in the region of 11 MeV. 
Looking at the transition densities, there is no clear evidence for a well--defined isospin splitting going from lower-- to higher--energy peaks. 
However, by comparing isovector and isoscalar transition probabilities, one observes a strong isospin mixing and a suppression of the isoscalar strength in the higher--energy part of the shown distribution. This indicates the existence of an isospin splitting. 

The second part of this work was devoted to a qualitative estimation of BMF effects on the symmetry energy of infinite matter and its slope, starting from the computation of the percentage of the EWSR of the low--lying dipole spectrum (up to 12 MeV) with the SSRPA model. 
Mean--field--based calculations were used, with four Skyrme parametrizations, to visualize linear correlations between the percentage of the EWSR and $L$, $L$ and the neutron--skin thickness, as well as the dipole polarizability times the symmetry energy and the neutron--skin thickness. By performing linear fits on these points and computing associated uncertainty bands, we have estimated BMF effects on $L$, on the neutron--skin thickness, and on $J$. The mean--field values for $J$ and $L$ are 26.83 and 37.70 MeV, respectively, with the parametrization SGII. We have estimated for the same parametrization BMF values of $J=$ 27.617 $\pm$ 5.004 MeV and $L=$   
60.815 $ \pm$ 16.982 MeV. Both quantities are increased by BMF effects but it is clear that the slope of the symmetry energy is much more sensibly affected. This indicates that, qualitatively, BMF effects tend to lead to stiffer EOSs for pure neutron matter.

\begin{acknowledgments}
This project has received funding from the European Union Horizon 2020 research and innovation program under Grant No. 654002. 
\end{acknowledgments}

\end{document}